\documentclass[pre]{revtex4}
\usepackage{graphicx}
\usepackage{latexsym}
\usepackage{amsmath}
\usepackage{amssymb}
\usepackage{amsfonts}
\usepackage{amsthm}
\usepackage{epstopdf}

\begin{document}
\title{Novel nonlinear wave equation: regulated rogue waves
and  accelerated soliton solutions}
\author{ Abhik Mukherjee}
%\footnote{abhikmukherjeesinp15@gmail.com}

\affiliation{Department of Mathematics and Applied Mathematics, University of Cape Town, Cape Town, SOUTH AFRICA}
%\footnote{ms.janaki@saha.ac.in}
%$^2$
\author{Anjan Kundu}
%\footnote{anjan.kundu@saha.ac.in}
%$^3$

\affiliation{Saha Institute of Nuclear Physics\\
 Kolkata, INDIA}

\begin{abstract} 
A new exactly solvable (1+1)-dimensional  complex nonlinear wave
  equation exhibiting rich analytic properties has been introduced.
 A  rogue wave (RW), localized in space-time like Peregrine RW  solution,
  though richer due to the presence of  free parameters is discovered.
This freedom allows to regulate amplitude and width of the RW
 as needed. The proposed equation allows also
an intriguing topology changing accelerated dark soliton solution in spite
of constant coefficients in the equation.

\end{abstract}
\maketitle

%\pacs{
%42.65.Tg,
%optical soliton, nonlinear waves 
%02.30.lk,
%integrable system
% 05.45.Yv,
%soliton ..,
%42.65.-k,
%nonlin optics
%42.81.Dp,
%prop soliton
%42.50.Md
%self-induced transparency 
%92.10.H+,
%ocean waves & oscillation,
%02.30.jr,
%PDE
%42.65.Sf,
%spacio-temporal dynamics
%11.10.Lm,
%nonlin nonloc FT
%47.35.Fg
%solitary wave in Fluid mech}
\smallskip

% n solutions, modulation instabilities, exact lump soliton}
 % }
% %\submitto{\JPA}

%132
\section{Introduction}	
Upsurge of interest in exactly solvable nonlinear equation in (1+1)-
dimension is due to the exclusive properties of their solutions with
applicable potentials. Most important of such exact solutions is the
localized soliton solutions, which move with a constant velocity without any
change of their shape and size. Such equations can be real ones like Korteweg de Vries (KdV) equation,
modified KdV, Sine Gordon equations etc or complex like Nonlinear Schrodinger equation (NLS), DNLS, Hirota
 equations etc.

However in recent years, the emphasis has been shifted towards newer solutions
in diverse physical applicability, with attempts to discover new nonlinear
exactly solvable equations. Therefore, the focus has been concentrated
gradually on more intricate solutions like accelerating solitons \cite{accsol},
topologically nontrivial solutions\cite{toposol}, RW solutions etc \cite{Rogue}.

Though RWs are deep ocean surface waves with manifestly (2+1) dimensional 
properties \cite{ourpaper}, the concept of RW has been penetrated recently to 
(1+1) dimensions with a large number of experimental and theoretical
investigations in a wide range of fields \cite{Rogue}. The most
popular of such RW solutions is Peregrine breather solution. Such RWs unlike
usual soliton solutions grow in amplitudes while moving reaching to its
maximum at certain time and disappearing again to the background waves.
However in spite  of its generic nature and success in explaining RW
phenomena in experimental observations in diverse fields \cite{Bonatto,Solli,Onorato,Montina,Kibler,
Shats,Chabchoub}, the
description of RW by Peregrine breather is limited to a fixed maximum
amplitude (3 times of background wave) and fixed RW width. This is due to
the absence of any free parameters in Peregrine RW solutions \cite{Rogue}.

Most of the subsequent findings in the systems like Kundu Ekkaus equation \cite{kunduekkaus},
Davey Stewartson equation\cite{DS}, Hirota equation\cite{Hirota} etc are also devoid of free parameters in
single peak RW solution. Only in higher RW solutions with multiple peaks
which are physically less important such free parameters can arise \cite{HPB1,HPB2,HPB3}.
Therefore a natural question is to search for possible one peaked RW solution
for different nonlinear equation which would allow free parameters to
regulate the amplitude and width to fit different types of such waves
obtained in nature and real experiments.

The accelerating solitons, which is the focus of another current interest
arise on the other hand in inhomogeneous media, modelled by extending the
known integrable equations to their inhomogeneous forms by
introducing variable dependent coefficients. However, the logical curiosity
is whether accelerating solitons could appear even in a homogeneous medium
described by equation with constant coefficient. This would not only enhance
our understanding of the basic cause behind a variable solitonic velocity
and its shape but also could open up new applicability.
 A continued
interest in nonlinear physics is a fascinating world of topological soliton
solutions. Though in one dimension, its impact is not that prominent with
the appearance of stable kink solutions in Sine Gordon and $\phi^4$ theory
and dark solitons in NLS equation, undoubtedly have drawn much attention.
As we know, for obtaining kink and dark solitons and their multisoliton
variants one has to use different functional form of the solutions with
increasing complexity with the increase of soliton number. Therefore it
would be an intriguing question to investigate whether one can generate
the nature of multidark solitons using the simplest form of dark solitons,
given typically by $\tanh$ functions. 

Our present aim concerns  the above posed natural and inquisitive
questions, for investigating which we propose a novel exactly solvable 
 nonlinear complex wave
equation with current like nonlinearity.
%%%%%%%%%%%30/11/16
Within the framework of our nonlinear equation we investigate the
possibility of finding controllable one peak RW solution, accelerating and
topology changing dark solitons.

%%%%%%%%%%%%%%%%%%%%%%%%%%%%%%

{\it Proposed complex nonlinear wave equation :}
We introduce a novel complex wave equation 
\begin{equation}
 \frac{\partial^2}{\partial T^2}\psi - \frac{\partial^2}{\partial X^2}\psi =
2i \psi(J_X + J_T) ,
\label{oureqn1}
\end{equation}
along with its complex conjugate,
Where $J_X =  \psi^* \frac{\partial
\psi}{\partial X}-\psi \frac{\partial \psi^*}{\partial X}$ and $ J_T=
\psi^* \frac{\partial
\psi}{\partial T}-\psi \frac{\partial \psi^*}{\partial T}
 $ are current like nonlinearities. This equation in the lightcone coordinates 
($x= (T+X)/2, t =  (T-X)/2$) takes the simpler form
\begin{equation}
 \psi_{xt} + 2i\psi(\psi \psi_{x}^* - \psi^*\psi_{x}) = 0, 
\label{oureqn2}
\end{equation}
where the subscripts denote partial derivatives.
We show that, the equation (\ref{oureqn2}) allows a richer set of exact
solutions related to some important questions posed above.
It is noticeable that (\ref{oureqn1}) is a dispersionless equation and 
 allows nontrivial boundary condition at
space infinities
($|\psi| \to const.,\ \ $ $|x| \to \infty$), which will be significant for our
investigation. We use both forms (\ref{oureqn1}) and (\ref{oureqn2}) of our
equation whichever is convenient. We focus below on richer RW and dark
soliton solutions of the proposed exactly solvable nonlinear PDE. 
%%%%%%%%%%%%%%%%%%%%%%%%%%%%%%%%
%For this we first convert this equation to two real coupled equations by
%denoting the complex field $\psi(x,t)$ as $\psi = \rho e^{i \theta}$:  
%%%%%%%%%%%%%%%%%%%%%%%%%%%%%%%%%%%%%%%%%%%%%
%01/12/16
%%%%%%%%%%%%%%%%%%%%%%%

\section{Controllable rogue wave solutions:} 
\ \
Such sudden and  steep isolated waves with a single peak, borrowing the
concept  from the deep oceanic waves,
 are causing upsurge of theoretical and experimental studies in various
fields \cite{Rogue}-\cite{Chabchoub}. Such phenomena can be modelled by RW solutions defined
in (1+1) space-time dimensions.  
\begin{figure}[!h]
\centering
%\subfigure
{
 \includegraphics[width= 7cm, angle=-90]{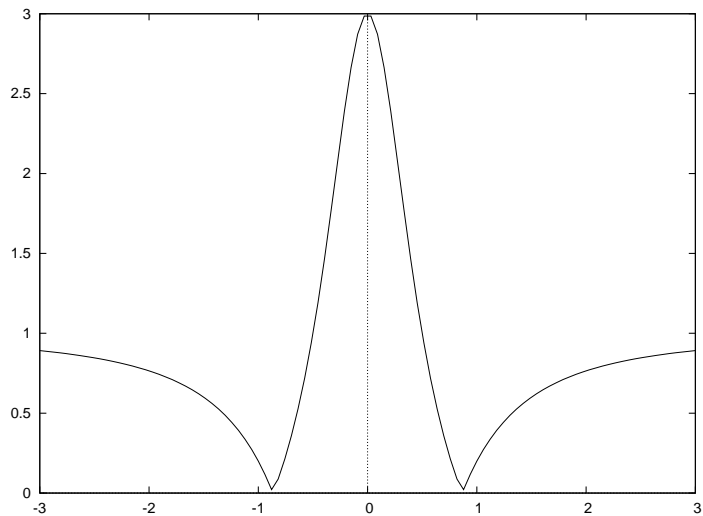}
}

\caption{ Amplitude variation of the full grown 1D rogue wave,
  modelled by the  modulus of the 
static Peregrine breather  .
 The maximum amplitude $3$ is
  attained at $x=0$,  while  it 
goes to  its asymptotic  value $1$  at $x \to \pm \infty $. The  maximum 
inclination attainable is  $3 \sqrt 3$ at $x=\frac{\sqrt{3}}{6}$,
 and becomes $0$ both at
$x=0$ and   $x \to \pm \infty $.
}
\label{fig:1}
\end{figure}

Therefore, Peregrine breather which offers such a simple solution (see Fig .1)
became the most popular RW model. However, as we see from the figure the
solution allows only fixed maximum amplitude, shape and width for the RW.
Since, in diverse situations observed experimentally the single peak RW can appear in
diverse amplitude, width and shape the application of Peregrine solution
becomes a bit restricted with a natural need for having richer RW solutions
allowing adjustable form of such solutions. Our aim here is to offer such a
general solution obtained from our proposed equation.

 Analyzing our nonlinear wave equation (\ref{oureqn1}) and with the physical
input that at space and time infinities the RW solutions
would go to linear background waves $e^{-2i(X-T)}$, while its amplitude will
grow nonlinearly and reach its maximum at the centre, we can extract the exact RW
solution in the explicit form 
\begin{equation}
 \psi = e^{-2i(X-T)}[-1 + \frac{1+2i(X-T)}{c+ (T-X)^2 + \frac{\alpha}{4}(X+T)^2 }],
\label{rog}\end{equation}
It is important to note that, the above solution includes two arbitrary
parameters $\alpha$ and $c$, which will be crucial for controlling the
physical
properties of this RW solution by changing these parameters as desired.
Looking closely to  solution (\ref{rog}), we see that at space-time infinities
$|X|\to \infty, \ \ |T|\to \infty$ the solution goes to the background traveling wave
and at  $T=X =0$, the amplitude attains its maximum value (
$\frac{1}{c} - 1$). It is evident therefore that the maximum amplitude of the RW
solution can be increased or decreased by varying the constant c as
demonstrated in Fig 2. This is a significant positive difference from the
Peregrine RW which shows the maximum amplitude fixed at $3$. Similarly it
reveals also that the width, inclination and shape of our RW solution
(\ref{rog}) becomes functions of two arbitrary parameters $\alpha$ and $c$
allowing us to control the property as needed to fit the experimental
observation. We mention again that such a tuning cannot be done through the
well known one-peaked Peregrine RW solution ( Fig.1), due to the absence of
free parameters.

\begin{figure}
\includegraphics[width=5cm]{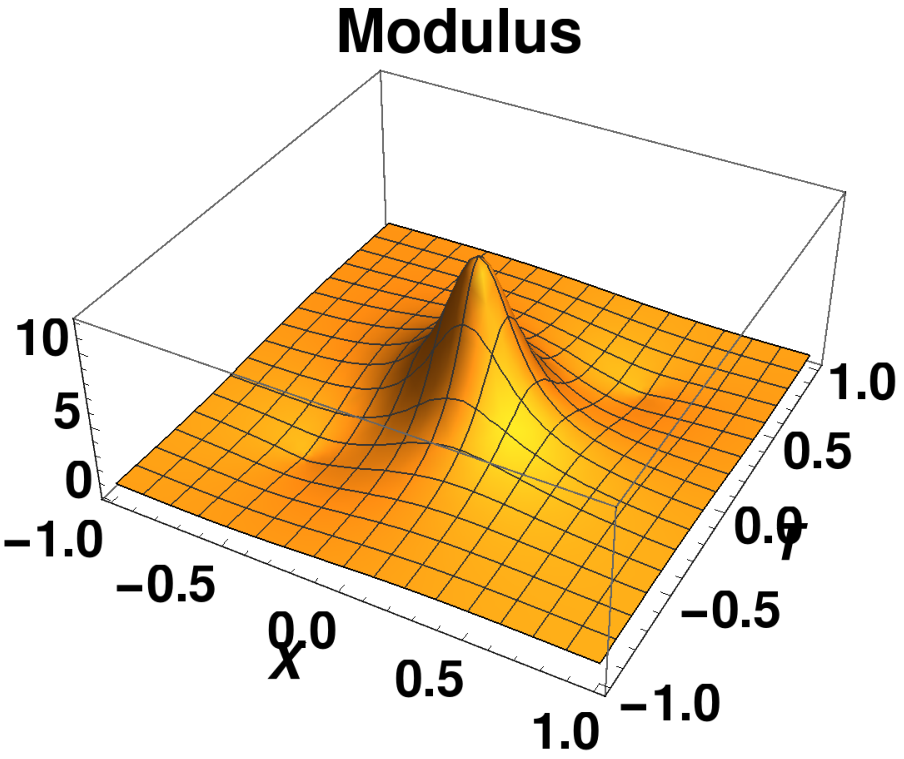}
 \ \ \ \includegraphics[width=5cm]{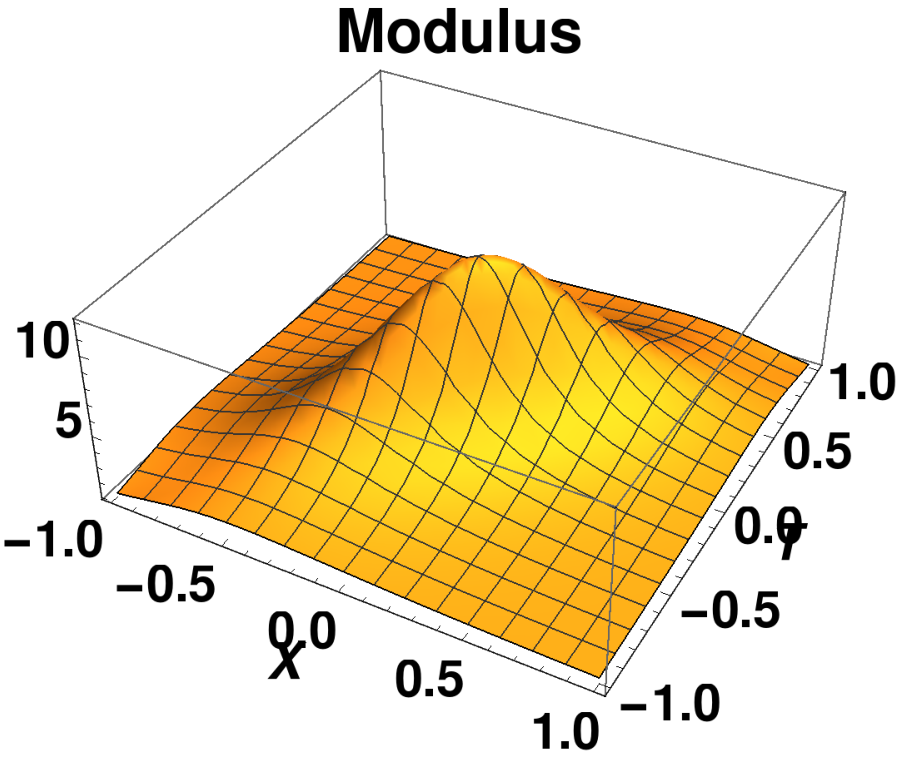}

 (a) 
\quad \qquad \qquad \quad \qquad (b)

\vspace{1cm}

\includegraphics[width=5cm]{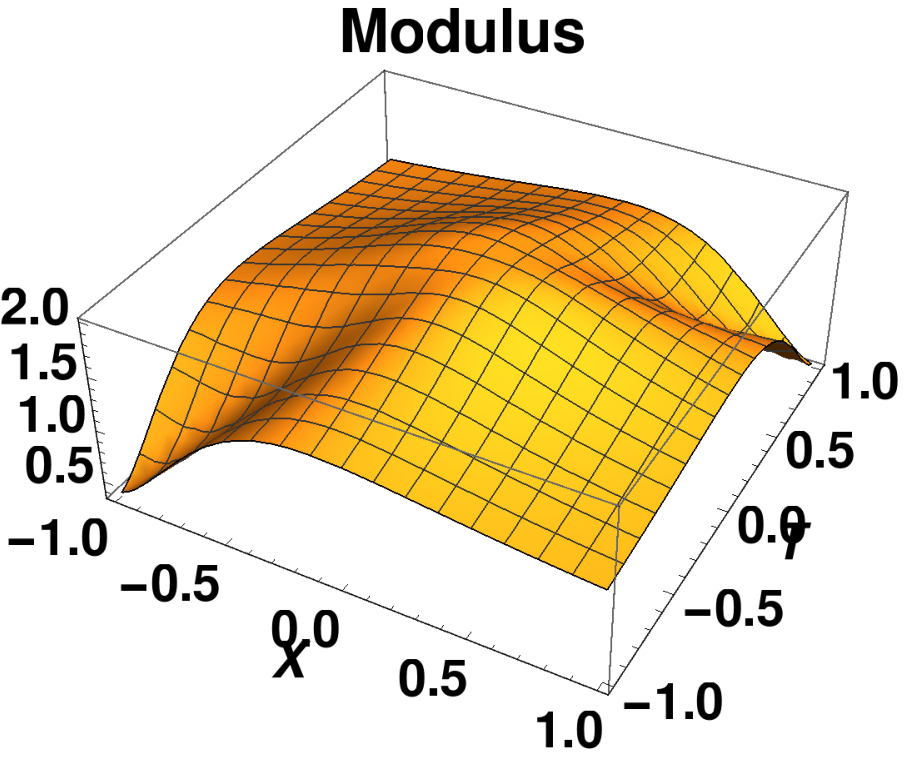}
 \ \ \ \includegraphics[width=5cm]{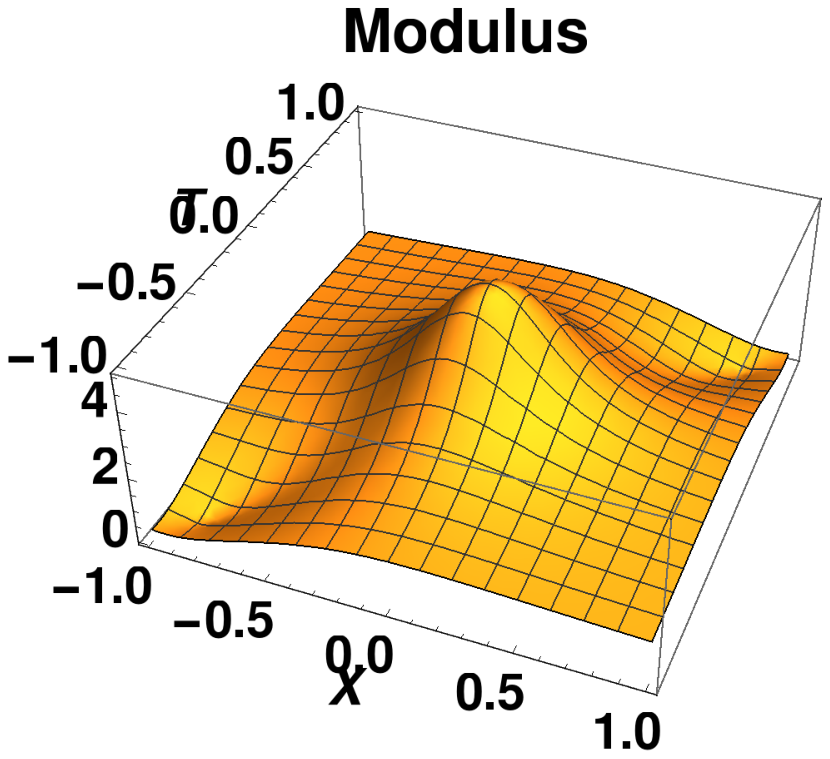}

\vspace{1cm}

 (c)
\quad \qquad \qquad  (d)

\noindent FIG.1: {Rogue wave (RW) modelled by the modulus of our solution (\ref{rog}) with different shapes
and sizes, generated from the same single peak solution. (a)High amplitude and high
steepness are obtained for $\alpha=4, c=\frac{1}{13}$.
(b)High amplitude and low
steepness are obtained for $\alpha=0.4, c=\frac{1}{13}$.
(c)Low amplitude and low
steepness are obtained for $\alpha=0.4, c=\frac{1}{3}$.
(d)Moderate amplitude and moderate
steepness are obtained for $\alpha=1.2, c=\frac{1}{6}$}.
\end{figure}

{\it Condition favorable for rogue wave solutions: modulation instability:}

After finding explicit RW solution in the proposed nonlinear equation
 we would like to investigate the condition crucial for the sudden generation of rogue wave
solution in our system out of the linear background waves, which is instigated mostly by
the modulation instability suffered by the wave due to the presence of
nonlinearity and dispersion in the media. This is a specific condition
between the frequency and wave vector, when
a small perturbation of the linear wave becomes unstable with a sudden
growth in time. Such a modulation instability is supposed to be crucial in
the formation of rogue wave in the physical system.

For easier calculations, we concentrate on the equivalent form (\ref{oureqn2})
for deriving the instability condition, known as modulation instability for
the  background wave, taking it as a perturbed linear wave form:
$\psi =(A_0 + \epsilon(x,t)) e^{i(\omega t + k x)}$. 
Inserting it in the nonlinear equation (\ref{oureqn2}) and neglecting higher
powers of $\epsilon$, we can derive a linear homogeneous equation for
$\epsilon$. Using the standard analysis, we can derive from this linear
equation, the condition for the modulation instability suffered by such a
perturbation given by the imaginary part of the modulation frequency which
in our case is $\omega_m = \pm{(i\sqrt{\omega^2 -4 A_0^2 \omega})}$.

Note that, this physical property of instability would retain the same for our
original nonlinear equation (\ref{oureqn1}) though its explicit form will
change due to change of the coordinate system.    

%%%%%%%%%%%%%%%%%%%%%%%%%%%%%%%%%% New formulae

%%%%%%%%%%%%%%%%%%%%%%%%%%%%%%%%%%%%%%

%%%%%%%%%%%%%%%%%%%%%%%%%%%%%%%%%%%%%%%%%%%%%
%%%%%%%%%%%%%%%%%%%%%%%%%%%%%%%%%%%%%%%%%

 \section{Accelerated dark soliton solutions:}

Unlike the RW solution the soliton solutions including the kink and the dark
solitons which propagates with constant velocity and amplitude without
change in their shapes are more prevalent in exactly solvable nonlinear
system. Therefore, our intension here is to look for the soliton
solutions in the simplest form $\psi(x,t)= \rho(x,t)e^{i \theta(x,t)} $ and
explore the possibility of finding the accelerated soliton
in our proposed equation which would be rather an intriguing problem
due to the homogeneity of the equation with constant coefficients. Inserting 
$\psi$ in (\ref{oureqn2}) and equating real and imaginary parts, we get two
coupled nonlinear equations:                                                 
\begin{equation}
\rho_{xt}- \rho \theta_{x}\theta_{t}+ 4 \rho^3 \theta_{x}=0,
\label{re} 
\end{equation}
\begin{equation}
 \rho_{x}\theta_{t} + \rho_{t}\theta_{x} + \rho \theta_{xt} = 0,
\label{im}
\end{equation}
For usual soliton solutions, carrier waves are given through
the phase $\theta = kx + \omega t+ \theta_0$ with constant frequency $\omega$
and wave vector $k$. However, we explore the possibility of generalizing it
nonlinearly as

\begin{equation}
\theta(x,t) = 	k x  + q(x) +( \omega t+p(t))+\theta_0.
\label{theta}
\end{equation}. 
where $q(x), p(t)$ are arbitrary  functions. Using the form (\ref{theta}), the
second equation (\ref{im}) gives

\begin{equation}
 (\omega +p^{'}(t))\rho_{x} = - (k + q^{'}(x))\rho_{t}
\label{rtx}
\end{equation}
which clearly allows the functional dependence of $\rho(\xi)$, with 
$\xi=\xi(x,t)$, giving 
\begin{equation}
 (\omega +p^{'}(t)) \rho_{\xi} \frac{\partial \xi}{\partial x} = - (k + q^{'}(x))\rho_{\xi}\frac{\partial \xi}{\partial t}
\end{equation}
Consistency gives the solution,
 %Let,$\frac{d\xi}{dt} = \beta_1$ and $\frac{d\xi}{dx} = -(\alpha_1 + q_{x})$
$\xi = \nu (k x + q(x) - (\omega t+p(t))+  \xi_0)$ keeping the same x dependence as in
$\theta$ but negating its t dependence, with $\nu, \xi_0 $ constants, which
may be chosen as desired. It is crucial to note here, that the appearance 
of the arbitrary functions in the argument $\xi$ in the present case,
 which will be responsible
for creating accelerating solitons, is an unusual property for a solitonic
equations. Notice that,  equation (\ref{rtx}) coming from the imaginary part
of our equation
contains only derivatives of function $\rho$ which can be traced back to the
appearance of arbitrary functions in $\xi$. On the other hand,
 the corresponding equation in the
wellknown NLS case contains function $\rho$, together with its derivatives,
which spoils this property and hence does not allow accelerating soliton in the
standard NLS equation.
%%%%%%%%%%% 08/12/16

The real part of the equation (\ref{re}) is simplified now using the 
explicit forms of $\theta$ and $\xi$ as 
\begin{equation}
 \nu^2 \rho_{\xi \xi} + \rho - \frac{4}{\omega+ p^{'}(t)} \rho^3 =0
\label{re1}
\end{equation}
Note that, the exact solution of the above equation does not allow any
function $p(t)$ other than constant. Therefore, considering the same
we look for explicit solutions of (\ref{re1}). 

On the other hand, for $\nu^2>0$, i.e, with real $\nu$ linked with the soliton
width, equation (\ref{re1}) with $p(t)=const.$ turns into the canonical form admitting
$\tanh$ solution representing dark soliton
\begin{eqnarray}
\rho = \frac{\sqrt{\omega}}{2} \ \ \tanh[ \xi],
 \ \xi = \nu (k x + q(x) - \omega t+  \xi_0).
\label{tanh}, \nu =\pm \frac{1}{\sqrt{2}}
\end{eqnarray}
 The appearance of the
arbitrary function $q(x)$ is a unique feature of our equation in spite of
its constant coefficients which differs drastically from the usual NLS and
derivative NLS families of equations. As mentioned above, the systems with
inhomogeneities modelled by x-dependent functions appearing explicitly in
the defining equations might show such a property of accelerated solitons 
 found here. We show interestingly, that this
freedom of choice of function $p(x)$ gives us in one hand an accelerating
dark soliton and on the other hand the solution with changing topological
properties.

For identifying the acceleration of soliton  (\ref{tanh}), we may
track the centre of the solution at $\xi =0$. Through simple steps, we may
derive the variable velocity $v(x)$ as well as the acceleration $a(x)$ as 
\begin{eqnarray}
v(x) = \frac{\omega}{k+ q^{'}(x)}, \ \ a(x) =-q^{''}(x)\frac{\omega^2}{(k +q^{'}(x))^3} 
\label{va}
\end{eqnarray}
 Obviously for $q(x)= x$, we get constant velocity solution,
however for any other choice of function $q(x)$ as the structure (\ref{va}) of the
velocity $v(x)$ and acceleration $a(x)$ suggests
that the soliton moves with a retarded velocity with the increase or decrease of
$x$. 

At the same time, the appearance of the arbitrary function q(x), the dark
soliton solution has an intriguing effect in the topological characteristic
of such soliton solutions defined through the boundary condition
as $Q = (\rho(x\to \infty) - \rho(x\to -\infty))$. For $q(x) =
const.$ we get a typical kink or antikink solution  
$\rho = \pm \frac{\sqrt{\omega}}{2} \ \ \tanh[\nu (k x  - \omega t]$ linked to 
$Q = \pm 1$, extrapolating between different vaccume.
 It is remarkable that, from other nonlinear choice for $q(x)$,
we start getting multi kink, antikink solutions allowing change of
topological properties generated from the same $tanh$ solutions (\ref{tanh})
which is atypical in soliton theory as well as for topological properties.
The following simple examples will illustrate the procedure, shown also in
Fig.3.

\noindent {\it Case} $q(x)=x^2$: It is easy to see that, in this case the
topological characteristic becomes trivial since 
$Q = \frac{1}{2}(\rho(x\to \infty) - (\rho(x\to
-\infty)=\frac{1}{2}(1 - 1)=0 $ which is due to the simultaneous appearance
of kink and antikink
extrapolating finally from the same vacuum. Therefore, we have created 2-
solitons (kink and antikink) from the same solution (\ref{tanh}). Recall
that for standard NLS case, kink-antikink solutions get more complicated
expression than $\tanh$.
%%%%%%%%%%%%%%%%%%%%131216

\noindent {\it Case} $q(x)=x^3$: This would lead again to 
$Q = \frac{1}{2}(\rho(x\to \infty) - (\rho(x\to  
-\infty)=\pm \frac{1}{2}(1 -(- 1))= \pm 1 $, showing the  generation of  3-
solitons (kink, antikink and kink) from the same solution (\ref{tanh}).

\begin{figure}
\includegraphics[width=7cm]{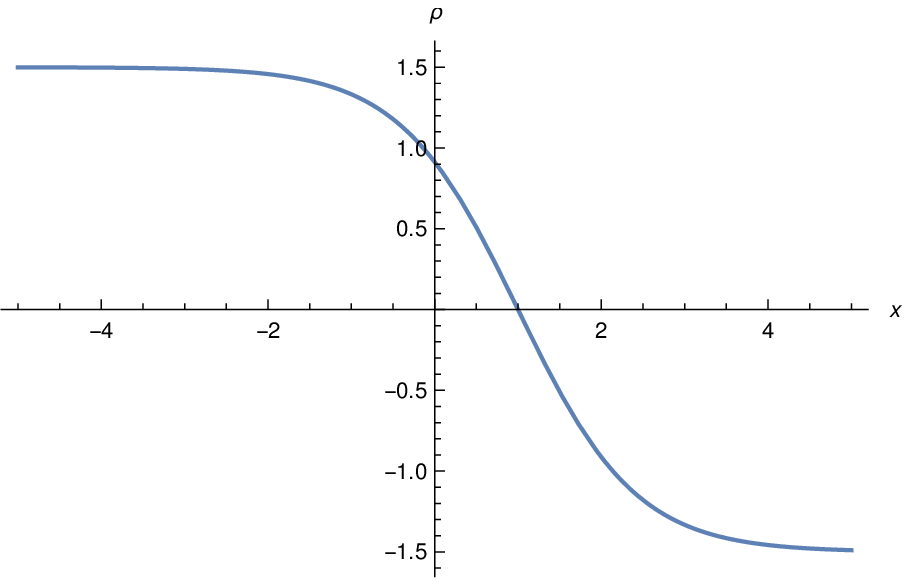}
 \ \ \ \includegraphics[width=7cm]{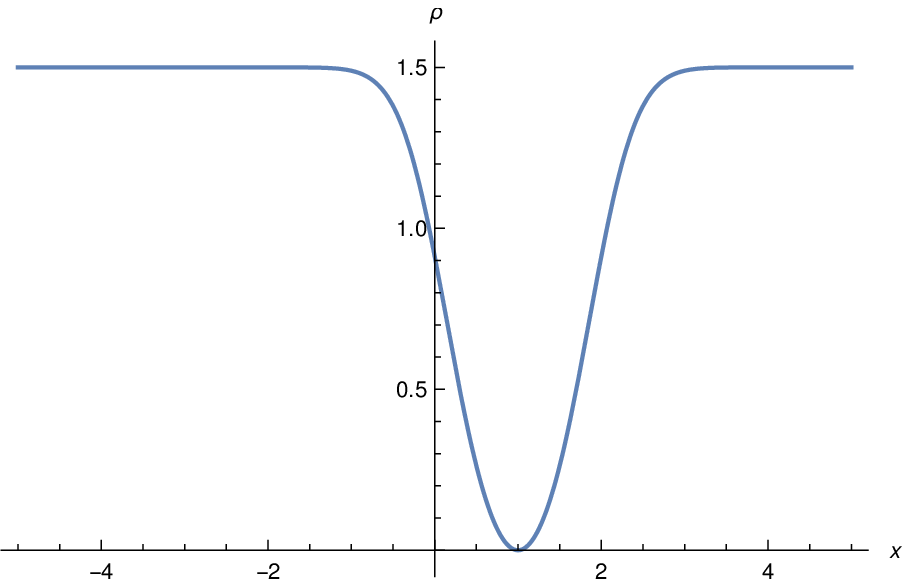}

 (a) $q(x) = x$
\quad \qquad \qquad \quad \qquad (b)$q(x) = x^2$

\vspace{1cm}

\includegraphics[width=7cm]{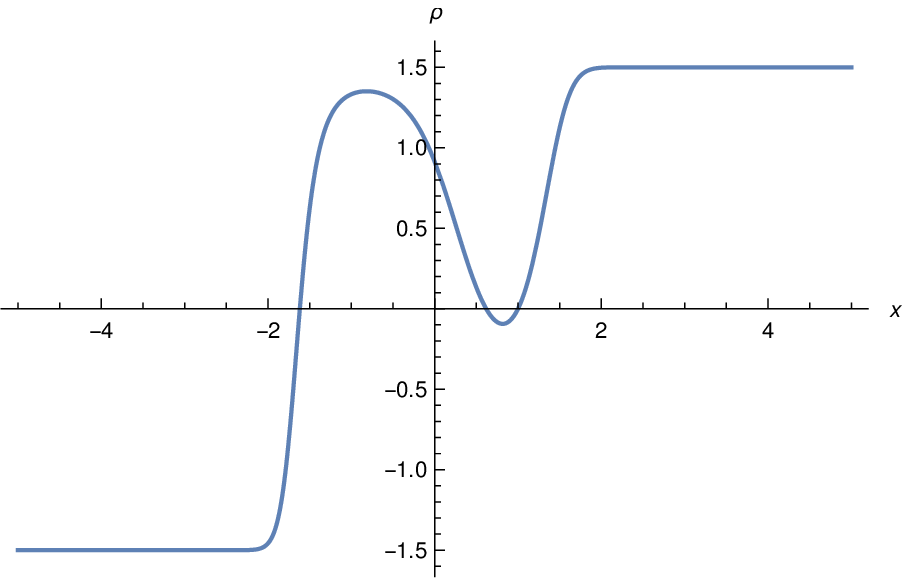}
% \ \ \ \includegraphics[width=7cm]{qx4.eps}

%\vspace{1cm}

 (c)$q(x) = x^3$
%\quad \qquad \qquad  (d) $q(x)=x^4$

\vspace{1cm}

\noindent FIG.2: {Accelerated soliton modelled by our solution (\ref{tanh}) at t=0, with different choices of arbitrary functions $q(x)$  and for the choice of constants
k = -2,
w = 3,
$\theta_0$ = 1;
$\xi_0 = 1$ }.
\end{figure}

 \section{Conclusive remarks:}

A new exactly solvable (1+1)-dimensional complex nonlinear wave equation has been introduced in this work. A rogue wave (RW) solution, localized in space-time 
and richer due to the presence of free parameters is discovered. This freedom allows to
regulate amplitude and width of the RW as needed. The proposed equation allows also an intriguing
topology changing accelerated dark soliton solution in spite of constant coefficients in the equation. This externally controllable solutions may have its
useful application in various branches of science and may pave new direction of research.

\section{Acknowledgement}

Abhik Mukherjee acknowledges his colleagues for valuable discussions.

\end{document}